\documentclass{sigchi}
\CopyrightYear{2016}
\setcopyright{acmcopyright}
\doi{http://dx.doi.org/10.475/123_4}
\isbn{123-4567-24-567/08/06}
\conferenceinfo{CHI'16,}{May 07--12, 2016, San Jose, CA, USA}
\acmPrice{\$15.00}

\usepackage{balance}       % to better equalize the last page
\usepackage{float}
\usepackage{graphics}      % for EPS, load graphicx instead 
\usepackage[T1]{fontenc}   % for umlauts and other diaeresis
\usepackage{txfonts}
\usepackage{mathptmx}
\usepackage[pdflang={en-US},pdftex]{hyperref}
\usepackage{color}
\usepackage{booktabs}
\usepackage{textcomp}

\usepackage{microtype}        % Improved Tracking and Kerning
\usepackage{ccicons}          % Cite your images correctly!
\usepackage{todonotes}
\def\plaintitle{SIGCHI Conference Proceedings Format}

\def\emptyauthor{}
\def\plainkeywords{Authors' choice; of terms; separated; by
  semicolons; include commas, within terms only; required.}

\makeatletter
\def\url@leostyle{%
  \@ifundefined{selectfont}{
    \def\UrlFont{\sf}
  }{
    \def\UrlFont{\small\bf\ttfamily}
  }}
\makeatother
\def\pprw{8.5in}
\def\pprh{11in}

\definecolor{linkColor}{RGB}{6,125,233}
\hypersetup{%
  pdftitle={\plaintitle},
  pdfauthor={\emptyauthor},
  pdfkeywords={\plainkeywords},
  pdfdisplaydoctitle=true, % For Accessibility
  bookmarksnumbered,
  pdfstartview={FitH},
  colorlinks,
  citecolor=black,
  filecolor=black,
  linkcolor=black,
  urlcolor=linkColor,
  breaklinks=true,
  hypertexnames=false
}

\begin{document}

%
% The "title" command has an optional parameter, allowing the author to define a "short title" to be used in page headers.
\title{Accessibility of Virtual Reality Locomotion Modalities to Adults and Minors}

%
% The "author" command and its associated commands are used to define the authors and their affiliations.
% Of note is the shared affiliation of the first two authors, and the "authornote" and "authornotemark" commands
% used to denote shared contribution to the research.
\numberofauthors{7}
\author{%
  \alignauthor{Zhijiong Huang\thanks{The first two authors contributed equally to this paper.} \\
    \affaddr{UC Berkeley}\\
    \affaddr{Berkeley, United States}\\
    \email{zhijiong@berkeley.edu}}\\
  \alignauthor{Yu Zhang\footnotemark[1]\\
    \affaddr{UC Berkeley}\\
    \affaddr{Berkeley, United States}\\
    \email{yu\_zhang@berkeley.edu}}\\
  \alignauthor{Kathryn C. Quigley\\
    \affaddr{Lawrence Hall of Science}\\
    \affaddr{Berkeley, United States}\\
    \email{kquigley@berkeley.edu}}\\
  \alignauthor{Ramya Sankar\\
    \affaddr{Lawrence Hall of Science}\\
    \affaddr{Berkeley, United States}\\
    \email{rsankar@berkeley.edu}}\\
  \alignauthor{Clemence Wormser\\
    \affaddr{UC Berkeley}\\
    \affaddr{Berkeley, United States}\\
    \email{clemence.wormser@berkeley.edu}}\\
      \alignauthor{Xinxin Mo\\
    \affaddr{UC Berkeley}\\
    \affaddr{Berkeley, United States}\\
    \email{xinxinmo@berkeley.edu}}\\
          \alignauthor{Allen Y. Yang\\
    \affaddr{UC Berkeley}\\
    \affaddr{Berkeley, United States}\\
    \email{yang@eecs.berkeley.edu}}\\
}

\maketitle

%
% By default, the full list of authors will be used in the page headers. Often, this list is too long, and will overlap
% other information printed in the page headers. This command allows the author to define a more concise list
% of authors' names for this purpose.

%
% The abstract is a short summary of the work to be presented in the article.
\begin{abstract}
Virtual reality (VR) is an important new technology that is fundamentally changing the way people experience entertainment and education content. Due to the fact that most currently available VR products are one size fits all, the accessibility of the content design and user interface design, even for healthy children is not well understood. It requires more research to ensure that children can have equally good user \hyphenation{experiences} compared to adults in VR. In our study, we seek to explore accessibility of locomotion in VR between healthy adults and minors along both objective and subjective dimensions. We performed a user experience experiment where subjects completed a simple task of moving and touching underwater animals in VR using one of four different locomotion modalities, as well as real-world walking without wearing VR headsets as the baseline. Our results show that physical body movement that mirrors real-world movement exclusively is the least preferred by both adults and minors. However, within the different modalities of controller assisted locomotion there are variations between adults and minors for preference and challenge levels. 
\end{abstract}

%\settopmatter{printacmref=false, printfolios=false}



%
% Keywords. The author(s) should pick words that accurately describe the work being
% presented. Separate the keywords with commas.

%
% A "teaser" image appears between the author and affiliation information and the body 
% of the document, and typically spans the page. 

%
% This command processes the author and affiliation and title information and builds
% the first part of the formatted document.

\section{Introduction} \label{sec:introduction}
Recent advancements in Virtual Reality (VR) have disrupted lots of traditional business, thereby changing many aspects of our daily lives. However, accessibility issues hinder people from adopting VR and restrict its application domain. In User Experience and Game Design literature, accessibility issues typically refer to how to design software and computers to make them more accessible to humans with various types of impairments.

In particular, there are three types of sub-problems that lead to accessibility issues: the constraint to receive feedback from the software due to a sensory impairment, the constraint to provide input to the software due to a motor impairment, and the constraint to not being able to understand the meaning of the software content due to a cognitive impairment.
In the emerging VR field, accessibility problems can be magnified to create bad experiences for users, mainly because the interactions are designed in 3D space that typically require the user to provide 6 degree-of-freedom (DOF) input and to perceive the 3D environment from their egocentric perspective. 

In this paper, we focus on understanding the accessibility issue of locomotion techniques in virtual reality for minors in comparison with adults. We identify five main modalities of locomotion in VR with wearable VR head-mounted display (HMD) and hand-held controller(s)
\begin{enumerate}
    \item Mapping human lower body movement to the movement of the egocentric perspective in VR, typically involving the walking motion.
    \item Mapping human upper body movement to the movement of the egocentric perspective in VR, typically involving the flying and swimming motion.
    \item Using buttons on the controller to provide direction and velocity commands, which resembles driving a vehicle in racing games.
    \item Teleportation, which refers to transporting the user's egocentric perspective instantaneously to a 3D location selected by the controller.
    \item Any combination of the above modalities.
\end{enumerate}

Our target subject population includes adults (age older than 17 years) and minors (age older than or equal to 7 and younger than or equal to 17 years) who do not have known physical and cognitive impairments. In this setting, we want to understand whether the different physical and cognitive capabilities between healthy adults and minors would lead to different user experience when the goal is to move their virtual egocentric perspective to achieve certain goals in VR environment. We believe better understanding of the question is significant in improving the UI/UX design in 3D VR games for both adults and minors. For example, the majority of the 3D VR experience available today are designed with a typical adult user in mind, in part because the current generation consumer VR devices have not been custom-built for young minors. In fact, several commercial VR systems specifically advise against usage by minors younger than 13 years old. When the software design does not consider the physical and cognitive variations from younger population, their user experience may suffer. This creates a barrier for adopting VR technology for minors in the education sector and other fields.

To address these issues, we set up an experiment in VR to investigate the validity of the following two hypotheses:
\begin{enumerate}
    \item In VR, using controller assisted movement modalities can improve the user experience compared to that using only physical body movements.
    \item There exist variations in the challenge level and preference using these locomotion modalities between adults and minors.
\end{enumerate}

In order to test our hypotheses, we chose to design a simulated VR experience in an underwater environment. In an underwater environment, the human avatar has true 6 DOF in body movement, and the real-world motion would involve both the upper body and the lower body. Therefore, it is more challenging and complex to control underwater body movement than moving on the ground in a stand-up posture. We believe this experimental setup enables us to thoroughly investigate the accessibility issues of the locomotion modalities.

\section{RELATED WORK}
Previous academic works relevant to this study are divided into two categories. The first category is the research illustrating principles, benefits, and drawbacks of different locomotion techniques. The second is to understand challenge and preference variations between different age groups' for locomotion systems in VR.

According to \cite{1}, which provides a comprehensive overview of VR locomotion techniques, the locomotion techniques are classified into steering travel, selection-based travel, manipulation-based travel, and walking locomotion. For example, steering techniques include gaze-directed steering with which the user is moving in their gaze direction or optionally in lateral directions \cite{2}. Selection-based travel requires the user to perform a selection task, e.g., by pointing to a destination in the virtual world, which will teleport the user's viewpoint to the target location \cite{3,4,5}. Manipulation-based locomotion is motivated by user's body motion \cite{1}. For example, drawing a circle in the air makes the user move forward. Walking locomotion could be taken as a natural travel technique when the VR settings are based on the ground, which is implemented by mapping users' motion information in the virtual world with that in the real world \cite{6}. 

Additionally, some previous works compare user outcomes for VR locomotion techniques. The joystick technique, which is an instance of steering travel, has been reported to lead to a significant increase in motion sickness compared to the other locomotion techniques \cite{7}. Teleportation, which is an example in the selection-based travel category, requires some time for users to understand their new surroundings after teleporting. This potentially leads to disorientation and can break immersed feeling for users \cite{8,9,10}. Finally, natural locomotion techniques like walking have been reported to be slightly more advantageous than semi- or non-natural techniques when it comes to a sense of presence or user preference \cite{4,10}.

Another group of research focuses on comparing how different age groups walk in VR compared to walking in the real world. In one research study conducted by Omar et.al. \cite{11,12}, the spatiotemporal parameters for young adults (age 18-34 years) and older adults (age 45-83 years) in both reality and VR are measured. The results indicate that older adults have similar walking biomechanics in both conditions. On the other hand, for young adults the gait speed is slower and steps are shorter in VR compared to the real world \cite{11}. This result is also supported by \cite{12}, which had 19 participants with ages between 18-38.

With respect to the research in this paper, the natural locomotion implementation is defined as swimming motion instead of walking, because it is more natural for users to swim rather than walk in an underwater world. A group in the NHTV University of Applied Sciences Breda conducted research related to the swimming locomotion. They created a diving game, in which players swim with a virtual Diver Propulsion Vehicle. The paper demonstrated that minors have more consistency between the real world and VR than the adults \cite{13}, which could imply that people of different ages may have their own preference for swimming locomotion.

\section{Methodology}
In order to study the accessibility of different locomotion methodologies for healthy adults and minors with respect to the two hypotheses in Section \ref{sec:introduction}, we have chosen to implement four modalities in a simulated underwater VR environment, and designed an experiment to investigate the performance and preference for these modalities. 

\subsection{Four Locomotion Modalities}
  The four locomotion modalities and one baseline modality are described in Table \ref{locomotion}, one type Swimming is considered a natural locomotion based on physical movements, while the other three are controller assisted locomotion. The first technique \textit{Swimming} mirrors body movement when swimming in real life. The implementation of this modality is to detect the motion, velocity, and position of controllers controlled by the user's two hands. The user needs to pull the triggers at the beginning and release them after each stroke. Controllers' motion during each stroke is mapped to movement of the user's avatar. The faster the user moves the controllers the faster they moves in the VR, mirroring the natural swimming process. Steering is controlled by the starting and ending point of the controllers for each stroke. 
  
  Second, we use controllers to allow subjects to teleport. A user can move to a selected location instantaneously by pulling and releasing a trigger defined as the teleportation function on the controller. The location itself is determined by the orientation of the controller for the direction and then by touching a forward or backward trigger for choosing the distance. In this study, we classify this as a controller assisted locomotion technique. 
  
  Third, we define another controller assisted modality called \textit{Look \& Follow}. The modality is implemented by defining the direction of the user movement through the orientation of the user's head pose, which is typically tracked in real time by the HMD. It then defines the movement speed by pulling a trigger on the controller.
  
  Finally, we create an \textit{Assisted Swimming} modality as a combination of \textit{Swimming} and \textit{Look \& Follow}, where subjects would move the controller as in a swimming method, but pull the trigger to move with the same steering motion as the \textit{Look \& Follow} modality. We will not tell the users that moving the arms in swimming motion actually had no effect to the actual movement in VR. Instead, the actual movement was determined exactly as the \textit{Look \& Follow} modality. The purpose of adding a "fake" swimming motion is to enhance the immersiveness of the locomotion for the underwater environment. We also classify this as a controller assisted locomotion technique. 
  
  In order to compare the challenge level and preference of the four locomotion modalities, we have also created a baseline where the subjects are asked to walk naturally to complete similar tasks in a room without wearing VR devices. In our study the baseline serves as a point of comparison to see how challenge level and preference compare to the neutral, everyday experience of walking. 

\begin{table}[th!]
\caption{Four Locomotion Modalities in VR and a Baseline}
\label{locomotion}
 \begin{tabular}{|p{1.6cm}|p{5.5cm}|} 
 \hline
 Locomotion Modalities & Descriptions \\ [0.5ex] 
 \hline
Swimming & 
The subjects hold two controllers and mimic the motion of swimming in the real world. 
%(Although implemented through manipulation-based technique, but it can be seen as a natural travel approach in an underwater game setting.)
 \\ 
 \hline
Look \& Follow & 
Subjects will move in the direction they are facing when pulling the trigger on controller. 
%(steering travel)
  \\
 \hline
Teleportation & 
Subjects will see a target ball when pulling the trigger. They will traverse to the position of the targeted ball instantaneously after releasing the trigger. 
%(selection-based travel)
\\
 \hline
Assisted Swimming & 
Subjects follow a natural swimming motion with controllers, but use the trigger to move with the \textit{Look \& Follow} locomotion technique. This mode is a combination of the first two modalities. 
% i.e. natural travel plus steering travel.
  \\
 \hline
Baseline & 
Subjects walk around a room naturally to complete a simple task without wearing VR devices.
 \\ 
  \hline
\end{tabular}
\end{table}

\subsection{Tasks} 

In order to compare the challenge level and preference of the four modalities, we set up a simple task. The requirement for the task is that subjects need to move towards three different groups of fish and touch them in the VR space as three sub-tasks. At the beginning, the subject will always be placed in the same location relative to the three groups of marine animals, i.e., the origin of the world coordinate system. The average distance between the subject to each group of the animal is five meters. As the subject approaches any one group of the animals, if they can use one of the controllers to touch one of the animals within their arm's reach, the sub-task is deemed complete. Subsequently the subject can proceed to move towards the rest animal groups if those sub-tasks have not been complete.

To experience each locomotion modality, a subject would need to repeat the above task (that contains three separate sub-tasks) for each locomotion type. In our experiment, on average, it would take about 15 minutes for an adult or a minor to complete one task in VR, except in baseline where the duration is quite fast since it is natural to walk in the real world. 

This situation presents a problem when we enroll subjects who are minors, some of whom were as young as 7 years old. We have found that very few minors can complete more than one task in VR without feeling tired or distracted. Therefore very early on we can conclude that each subject, especially as a minor, can only be interviewed to experience one locomotion in VR. 

Under this limitation, we still need to frame a fair comparison to compare the performance and preference of the all four locomotion types. To achieve that, we rely on the use of the baseline as described in Table \ref{locomotion}. In particular, in each of the interview session, a subject regardless being an adult or a minor will be asked to complete one session of the baseline and one session of the one of the VR locomotion types. Then, quantitative and qualitative metrics that aim to measure the user experience will be collected, which we will describe next.
\begin{figure*}[h!]
\begin{center}
\includegraphics[width=\textwidth]{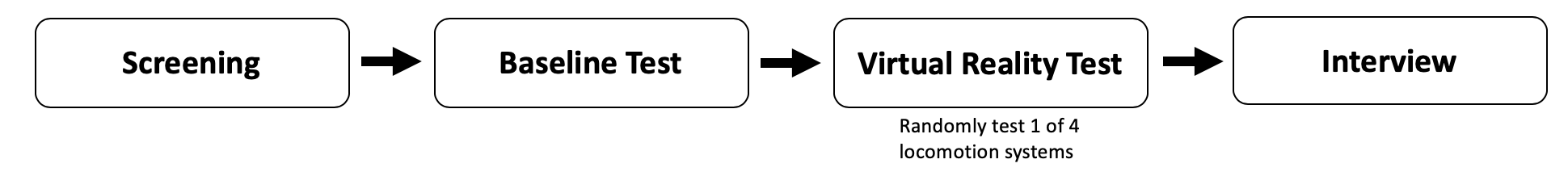}
\end{center}
   \caption{Flowchart for the Experiment Process}
\label{fig:short}
\end{figure*}

\subsection{Metrics}

We collected objective data for objective performance analysis, and collected subjective data for both challenge level and preference analysis. The objective metric for challenge level analysis was done by recording the completion time for each subjects experimental task in one of the four VR locomotion modalities.

\begin{table}[th!]
\caption{Questionnaire}
\label{Questionnaire}
 \begin{tabular}{|p{6.8cm}|p{1cm}|} 
  \hline
 Questions & Ranking \\[0.5ex]
 \hline
Q1. How would you rate the challenge level moving in real life on a scale of 0-10 (10 means most difficult)? & 0-10
 \\ [0.5ex]
 \hline
Q2. How would you rate the challenge level moving in virtual reality on a scale of 0-10 (10 means most difficult)?  &0-10
  \\[0.5ex]
 \hline
Q3. Comparing real movement with the virtual movement, which one do you find easier?  & Real or Virtual
\\[0.5ex]
 \hline
 Q4. How would you rate your preference level moving in real life on a scale of 0-10 (10 means most preferred)?&0-10
 \\
 \hline
 Q5. How would you rate your preference level moving in virtual reality on a scale of 0-10 (10 means most preferred)? &0-10
 \\
 \hline
 Q6. Comparing real movement with virtual movement, which one did you prefer? & Real or Virtual 
 \\
 \hline
\end{tabular}
\end{table}

For the collection of subjective data for both challenge level and preference analysis, we conducted short interviews with subjects, which is shown in Table \ref{Questionnaire}. For subjective challenge level evaluation we asked subjects to rate the difficulty level of the real-world baseline and locomotion modalities for a range between 0 and 10, see Q1 and Q2 in Table \ref{Questionnaire}. For the preference side, we asked subjects to rate their preference for both the baseline and one of the locomotion modalities for a range between 0 and 10, see Q4 and Q5 in Table \ref{Questionnaire}. Finally, in the interview with subjects Q3 and Q6 in Table \ref{Questionnaire} were set up as sanity checks for the subjects to make sure that they were answering consistently. 

Once we collected the data, our first step was to calculate a baseline for each dimension. Since the baseline was designed as a real-world experience, as shown in the Table \ref{locomotion}, we calculated mean values of all $40$ children subjects' assessment for the real-world experience for each dimension. In next step we obtained two mean values for each locomotion modality among the $10$ subjects who tried and gave evaluation to this that specific modality. The final step was the most important procedure, in which we compare the differences between the mean value of the locomotion modality and that of the baseline. The differences were regarded as relatively unbiased evaluation for each locomotion modality and the same process was applied on the data obtained from the adult group.
\subsection{Experiment Process Summary}

The flowchart in Figure \ref{fig:short} illustrates the experiment process. Before the experiment, subjects were asked to read and sign consent forms and a screening form. In the first step of the experiment, subjects were asked to complete a real-world task of walking in the room. Next, the subjects begin the VR portion of the experiment. 

In the experiment, each subject only tries one locomotion modality. First they do a short tutorial, which teaches the subject how to navigate in VR with the specific locomotion modality they are assigned with. The tutorial lasts until the subjects reach the target position in the tutorial or reach 80 seconds, whichever is shorter.

Once subjects are trained on how to move, they begin their the task mentioned above in VR with their assigned locomotion modality. The time limitation for the task is 300 seconds. We assigned the order for testing the four locomotion methods in advance to guarantee each modality would have metrics data from $10$ adults and $10$ minors. After the subjects completed both the real-world and VR tasks, they were invited to complete a short interview to assess their experience of the baseline and locomotion modalities in VR. 

\subsection{Selection of Participants}
A total of $100$ subjects were recruited randomly from people on campus and visitors to museum, comprising of $43$ adults aged between $19-65$ and $57$ minors aged between $7-17$. Each subject spent around $15$ minutes in the experiment and interview. The minors and adults were tested separately but following the exact same protocol. The Institutional Review Board (IRB) approval was obtained ahead of this experiment.

At the end of the experiment, $40$ sets of adult data and $40$ sets of minor data were considered valid. We rejected three sets of adult data and $17$ sets of minor data. There are two criteria that lead to the rejection of subject data. First, four minors who reported discomfort in VR and declined to complete the experiment. Second, $16$ sets of data were rejected due to subjects' inconsistent answers to the questions on the questionnaire (Table \ref{Questionnaire}). Q3 and Q6 were designed to be sanity checks and if subjects were inconsistent in an answer to Q3 or Q6, their data set was rejected. For instance, when asked about which experience they preferred (Q6 on the questionnaire) and to rate their preference (Q4 and Q5 on the questionnaire), some participants said they preferred the virtual experience to the baseline but gave a lower rate to the virtual experience.  

\section{Results}
In the study we calculated average of challenge and preference level for each type of locomotion. In addition we also calculated the average completion time for each type of locomotion for adults and minors. 

\subsection{Results from the Minor Group}
\begin{figure}[h!]
    \centering
    \includegraphics[width=0.45\textwidth]{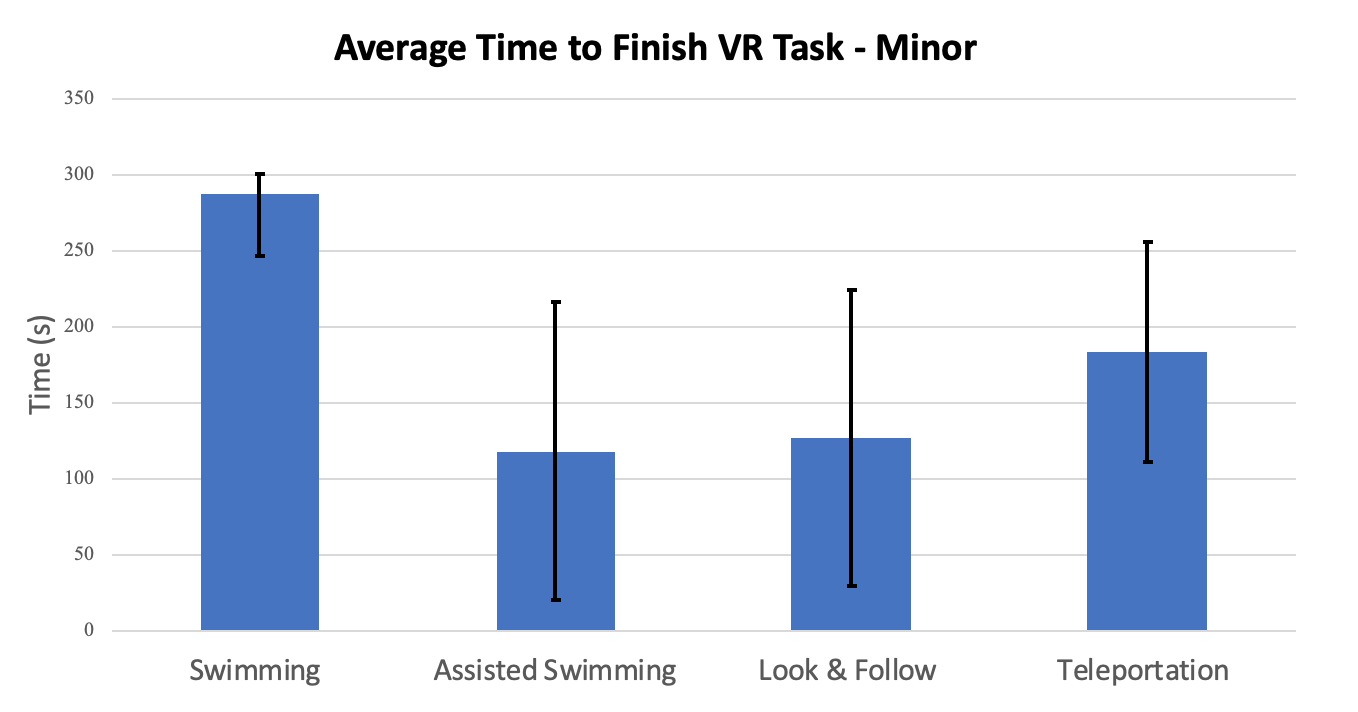}
    \caption{Average time and standard deviation for minors to finish a task in different modalities.}
    \label{Time_children}
\end{figure}

Figure \ref{Time_children} contains the average completion time and standard deviation (std) among minor subjects with respect to each locomotion modality. It shows that minors completed the task quickest with \textit{Assisted Swimming} modality ($avg. = 117.84s$) and slowest with the \textit{Swimming} modality ($avg. = 286.60s$).

\begin{figure}[h!]
    \centering
    \includegraphics[width=0.45\textwidth]{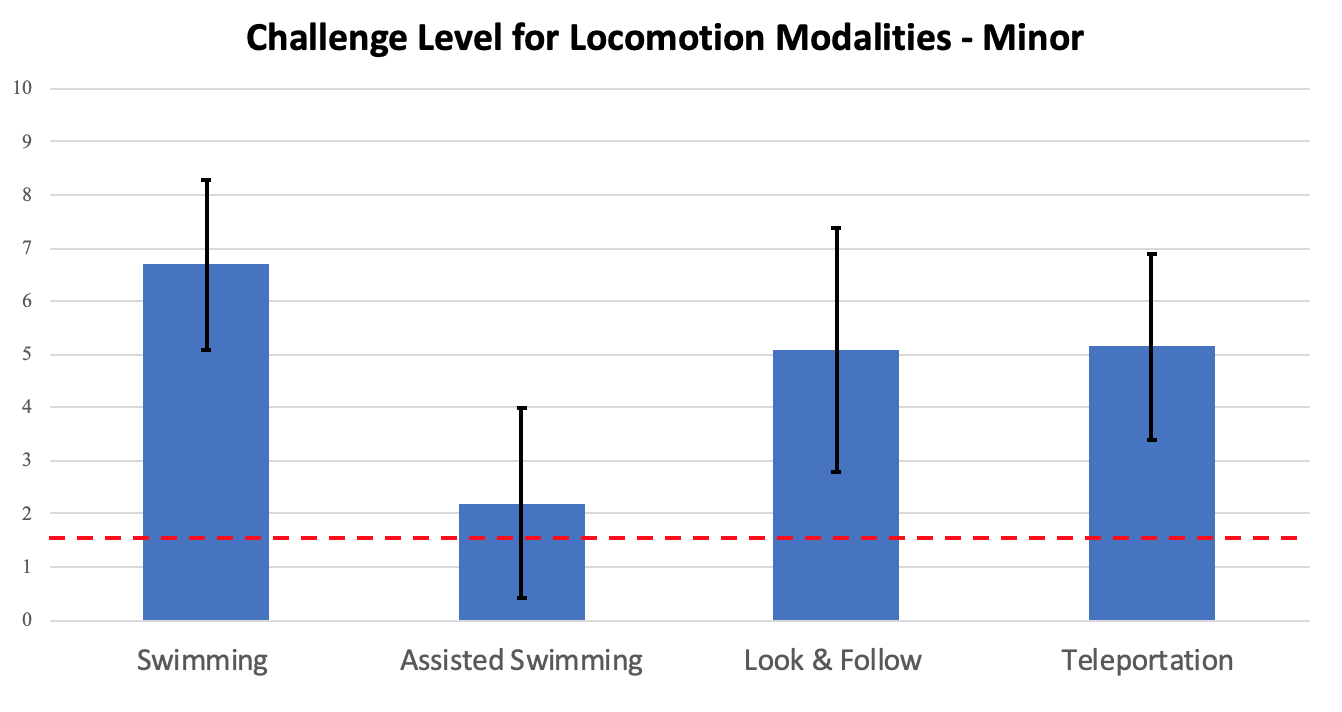}
    \caption{The average and standard deviation of the challenge level ranged between 0 to 10 from thehttps://www.overleaf.com/project/5c63abf57615f53d5f5beb56 questionnaire. The baseline numbers are average $1.68$ (shown as the red dash line) and standard deviation $1.59$.}
    \label{Challenge_children}
\end{figure}

Results of subjective assessment are presented in Figure \ref{Challenge_children}. The presented values are the mean values of each locomotion modality and the baseline, indicating that minors rated \textit{Assisted Swimming} with best average score ($2.10$) and rated \textit{Swimming} with worst average score ($6.70$). 

\begin{figure}[h!]
    \centering
    \includegraphics[width=0.45\textwidth]{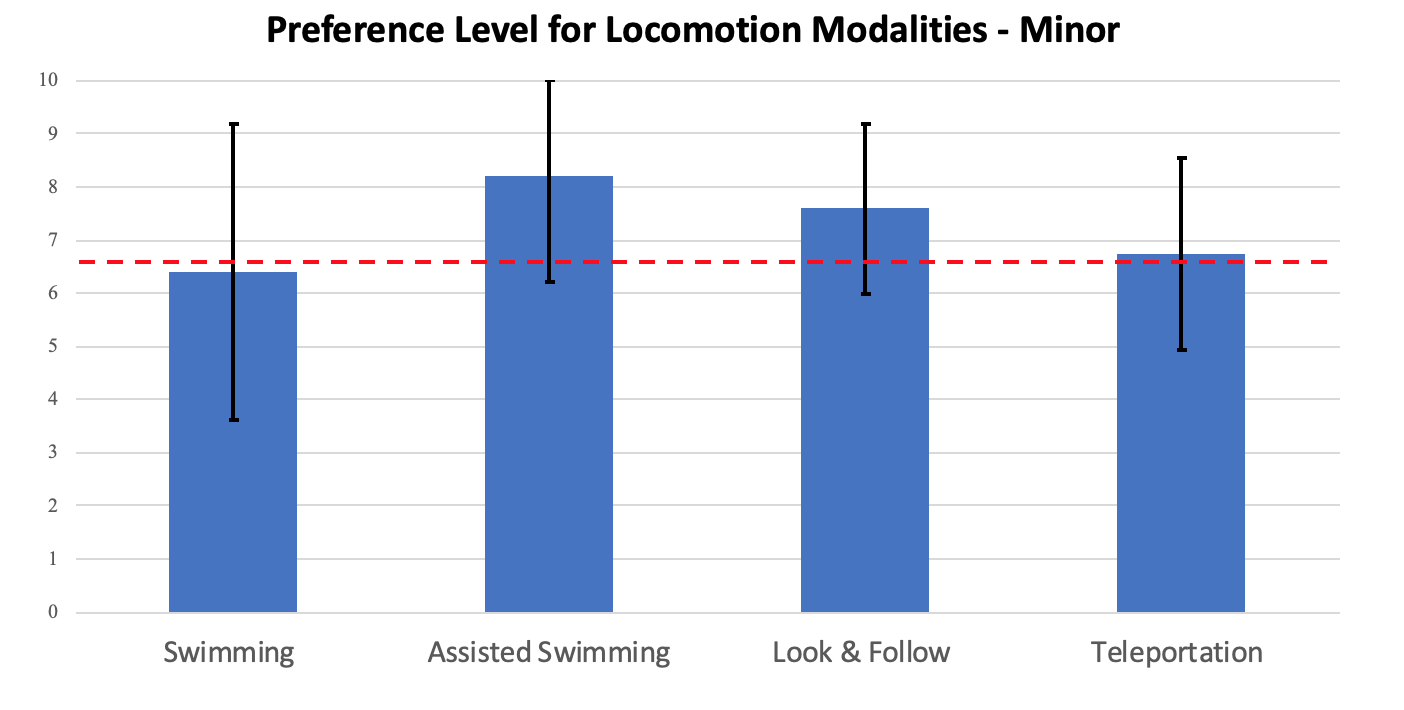}
    \caption{The average and standard deviation of the Preference level ranged between 0 to 10 from the questionnaire. The baseline numbers are average $6.63$ (shown as the red dash line) and standard deviation $2.29$.}
    \label{Preference_children}
\end{figure}

Figure \ref{Preference_children} illustrates the preference level of each modality, which is represented through the mean values of each locomotion modality and the baseline. The plot indicates that the \textit{Assisted Swimming} is minors' favorite locomotion modality ($avg. = 8.20$), and they rated the \textit{Swimming} with the lowest average score ($6.40$).

\subsection{Results from the Adult Group}
\begin{figure}[h!]
    \centering
    \includegraphics[width=0.45\textwidth]{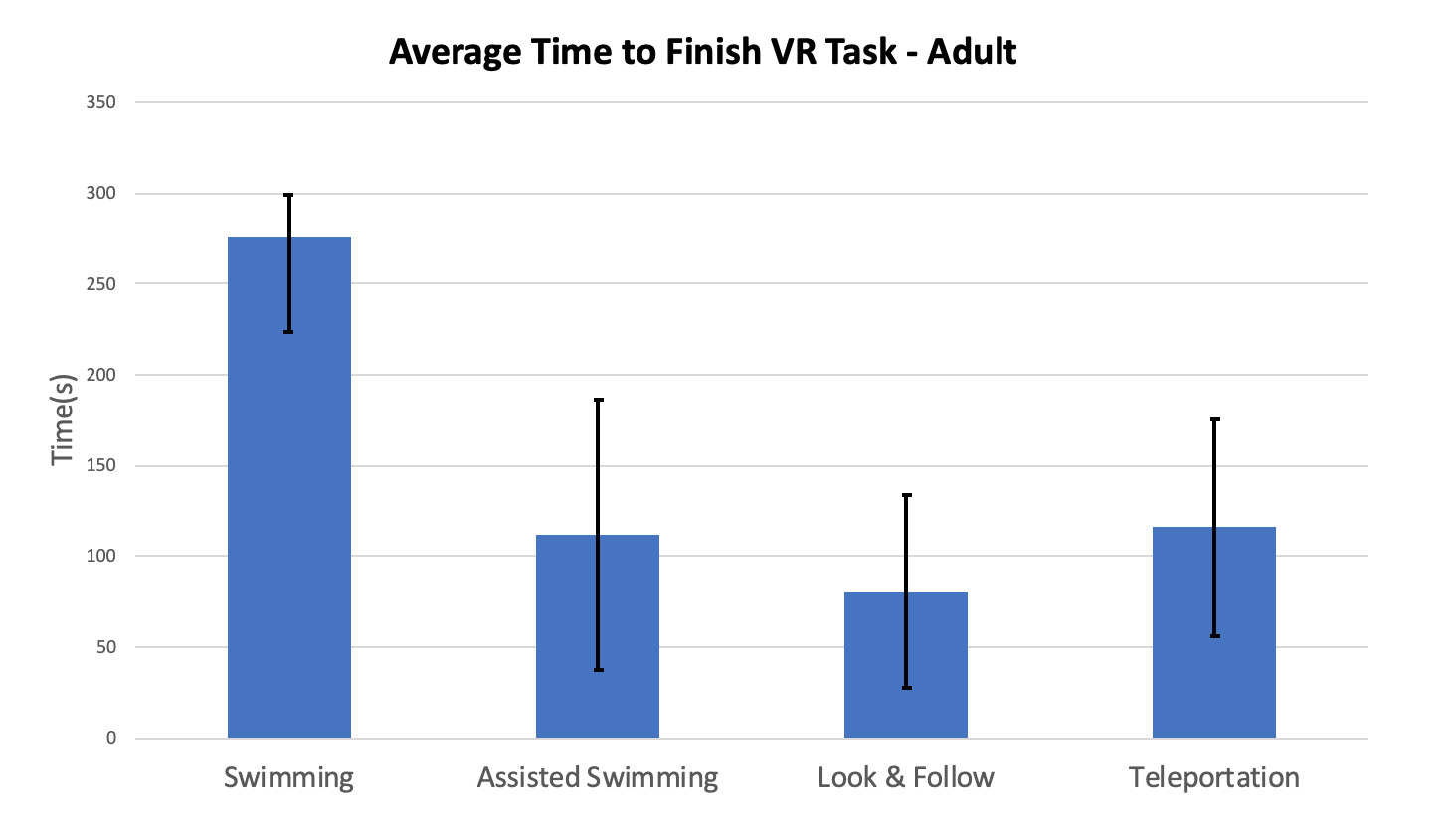}
    \caption{Average time and standard deviation for adults to finish a task in different modalities.}
    \label{Time_Adult}
\end{figure}

\begin{figure}[h!]
    \centering
    \includegraphics[width=0.45\textwidth]{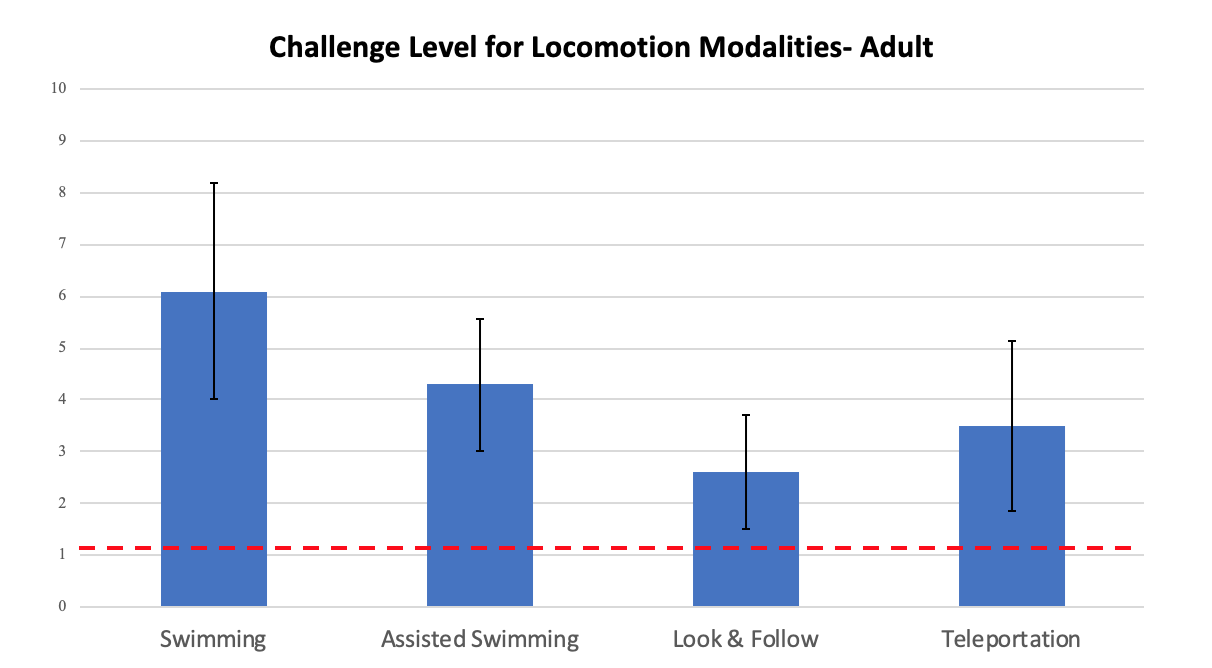}
    \caption{The average and standard deviation of the challenge level ranged between 0 to 10 from the questionnaire. The baseline numbers are average $1.15$ (shown as red dash line) and standard deviation $1.19$.}
    \label{Challenge_Adult}
\end{figure}

Figure \ref{Time_Adult} illustrates average completion time and std among adults subjects with respect to each locomotion modality. The graph shows that adults spent longest time completing the task with \textit{Swimming} modality ($avg. = 276.00s$), and adults completed tasks most quickly with \textit{Look \& Follow} ($avg. = 80.50s$). Furthermore, Figure \ref{Challenge_Adult} presents interviewee' subjective evaluation for each modality's challenge level. Adults rated \textit{Swimming} with the worst average score ($6.10$) and gave \textit{Look \& Follow} the best score ($2.60$).

\begin{figure}[h!]
    \centering
    \includegraphics[width=0.45\textwidth]{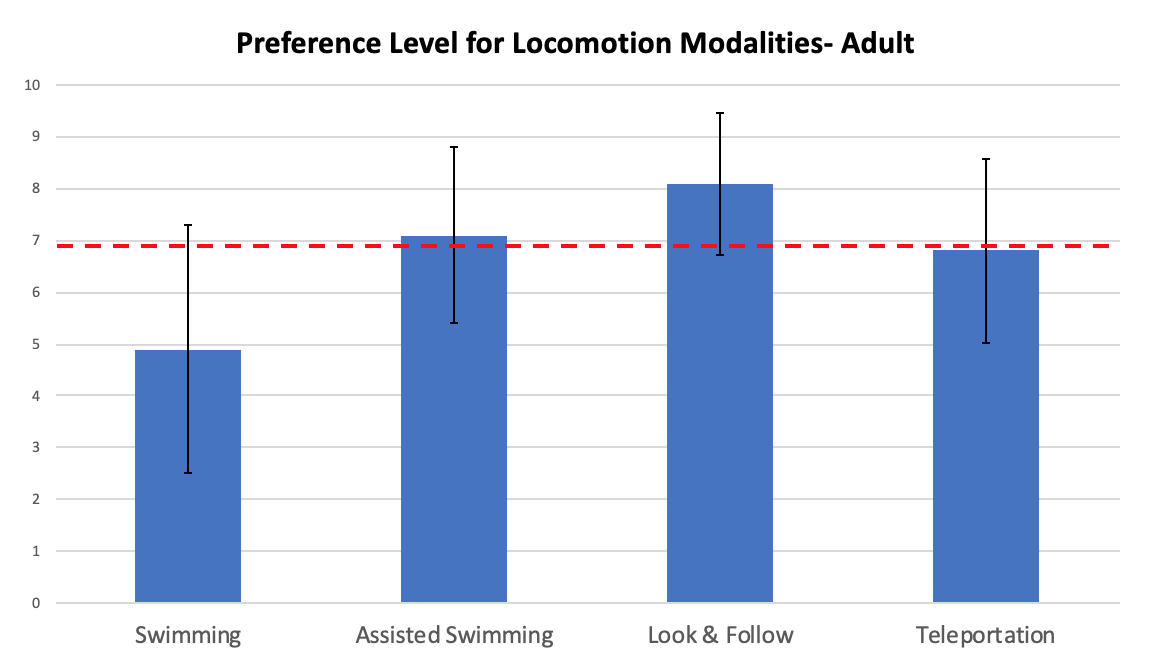}
    \caption{The average and standard deviation of the Preference level ranged between 0 to 10 from the questionnaire. The baseline numbers are average $6.93$ (shown as red dash line) and standard deviation $2.59$.}
    \label{Preference_Adult}
\end{figure}

As for the preference level, Figure \ref{Preference_Adult} illustrates adult group's evaluation, which is again represented by the mean values of the locomotion modality and the baseline. The plot indicates that adults liked \textit{Look \& Follow} the most with an average score of $8.10$, and they least preferred \textit{Swimming} with an average score of $4.90$. 

%\balance{}
\section{Discussion}
In our study we found similarities and differences between adults and minors for both challenge level and preference across the four locomotion modalities. 

\subsection{Accessibility Comparison Between Adults and Minors}
For the average time to finish the task, as presented in Figure \ref{Time_children} for minors and Figure \ref{Time_Adult} for adults, we found that minors spent least time completing the task with \textit{Assisted Swimming} modality ($avg. = 117.84s$) and \textit{Swimming} modality took them most time ($avg. = 286.60s$), indicating that the former being the easiest and the later being the most difficult for minors. Similarly, among adult group, the average completion time was highest for \textit{Swimming} modality ($avg. = 276.00s$), meaning that \textit{Swimming} modality is also the most difficult for adults. However, adults finished the task most quickly in the \textit{Look \& Follow} modality ($avg. =  80.50s$), which is different from that of minors, meaning that \textit{Look \& Follow} could be the easiest modality for adults. 

For the challenge level illustrated in Figure \ref{Challenge_children} for minors and Figure \ref{Challenge_Adult} for adults, we compared the results. The evaluation given by minors for \textit{Assisted Swimming} ($avg. = 2.1$) was quite close to their evaluation for the baseline ($avg. = 1.68$), while the other three modalities were significantly more challenging than the baseline. This means that minors regard \textit{Assisted Swimming} to be almost as easy as walking and the others much harder. Even considering the standard deviation for \textit{Swimming} ($std. = 1.60$) , \textit{Look \& Follow} ($std. = 1.63$), and \textit{Teleportation} ($std. = 1.74$), they are still higher than the baseline for walking ($std. = 1.59$), with \textit{Swimming} ranking as the most difficult at ($avg. = 6.70$). Adults assigned the lowest challenge level to the \textit{Look \& Follow} modality ($avg. = 2.6$), but still found it to be more difficult than the baseline for walking of ($avg. = 1.15$) even considering the standard deviation. The other three were much higher than the baseline, with \textit{Swimming} at ($avg. = 6.1$) ranked the most challenging modality. Additionally, compared to minors' group, results from adult group show relatively lower standard deviation in the evaluation, which may indicate more reliability in adults' data or possibly more variation in motor-skill abilities in minors.

The preference level between adults Figure \ref{Preference_Adult} and minors Figure \ref{Preference_children} showed some similarities and differences. We found that the adult preference higher than the baseline of ($avg. = 6.93$, $std. = 2.59$) was for \textit{Look \& Follow} at ($avg. = 8.1$, $std. = 1.37$), meaning that adults found the other three forms of locomotion either worse or the same as walking in the real-world. Although the adult preference for \textit{Assisted Swimming} at ($avg. = 7.1$, $std. = 1.7$) and \textit{Teleportation} ($avg. = 6.8$, $std. = 1.78$) are close to the adult preference baseline, the standard deviation makes those results uncertain. On the other hand, minors preferred both \textit{Assisted Swimming} at ($avg. = 8.2$, $std. = 2.00$) and \textit{Look \& Follow} at ($avg. = 7.3$, $std. = 1.50$) over the baseline of ($avg. = 6.63$, $std. = 2.29$), meaning they preferred these two movements over walking in the real-world, and the remaining two less than walking. Although minors' had a slight preference for \textit{Teleportation} of ($avg. = 6.75$, $std. = 1.81$) and \textit{Swimming} of ($avg. = 6.4$, $std. = 2.80$) near the baseline, due to the standard deviation these results were not certain.

\subsection{Hypotheses Verification}
Based on the discussion above-mentioned, we can verify the two hypotheses we have proposed.

First, both adult and minor groups show significantly more difficulty with \textit{Swimming} motion than the other three assisted motions, and subjects from both groups reflected obvious preference to the controller assisted locomotion over the natural \textit{Swimming} modality. Additionally, both adults and minors took much longer to complete the task using the natural swimming modality, which further indicates that it is more difficult to use. The finding confirms our first hypothesis, which states that using controller assisted movement modalities in VR can improve the user experience compared to locomotion which only uses physical body movements in VR. 

Secondly, although both groups found controller assisted locomotion more accessible, there is still variation for preference and challenge level within different types of assisted movement. For example, minors preferred \textit{Assisted Swimming} over walking and also found it easier than walking. This also correlates with \textit{Assisted Swimming} having the shortest average completion time for minors. Minors also had a preference for \textit{Look \& Follow} over walking, but found it more difficult than walking. On the other hand, adults found \textit{Look \& Follow} easiest and preferred it over walking, while finding all the others harder and less preferable to walking. Adults also had the shortest completion time for \textit{Look \& Follow}, supporting the result that it was top rated. These findings aligns well with our second hypothesis predicting the existence of variations between minors and adults.

\section{Conclusion}
In this paper, we conducted an experiment to evaluate the accessibility of virtual reality locomotion modalities to healthy adults and minors. We found that in virtual reality, assisted movement modalities could provide better experience to users than natural movement modality. On top of that, there are variations between adults and minors in the evaluation for each modality. 

In the following, we conjecture some reasons for the conclusions obtained above, which could be potential directions of future research on this topic.

The \textit{Assisted Swimming} modality turned out to be the most popular among minors, probably because it is less tiring than the \textit{Swimming} modality but at the same time is more immersive and more fun to do than \textit{Look \& Follow} and \textit{Teleportation}. 

Compared with minors, adults usually have better physiological conditions and many of them tend to be less sensitive to uncomfortable feelings such as locomotion sickness. Adults also have more life experience compared to minors, especially the experience of interacting with the environment. This difference could make it less attractive for adults to mimic swimming motions in virtual reality. Furthermore, compared to minors, many adults tend to value efficiency more, which means they may prefer methods that enable them to reach their goal without much effort. That may explain the adult group's preference for the \textit{Look \& Follow} mode over the \textit{Swimming} and \textit{Assisted Swimming} modalities.

Finally, if we look at the top two most preferred types of locomotion, they are between \textit{Assisted Swimming} and \textit{Look \& Follow}. If one would further consider the situation when subjects would have motor impairment to the upper or lower limps, which we did not consider in this paper, \textit{Look \& Follow} would be the most preferred modality and \textit{Assisted Swimming} would be less desirable.

\balance{}
\bibliographystyle{SIGCHI-Reference-Format}
\bibliography{ref}

%%%%%%%%%%%%%%%%%%%%%%%%%%%%%%%%%%%%%%%%%%%%%%%%%%%%%%%%%%%%%%%%%%%%%%%%%%%%%%%%

%%%%%%%%%%%%%%%%%%%%%%%%%%%%%%%%%%%%%%%%%%%%%%%%%%%%%%%%%%%%%%%%%%%%%%%%%%%%%%%%

%%%%%%%%%%%%%%%%%%%%%%%%%%%%%%%%%%%%%%%%%%%%%%%%%%%%%%%%%%%%%%%%%%%%%%%%%%%%%%%%

%%%%%%%%%%%%%%%%%%%%%%%%%%%%%%%%%%%%%%%%%%%%%%%%%%%%%%%%%%%%%%%%%%%%%%%%%%%%%%%%

%\balance{}
%\bibliographystyle{SIGCHI-Reference-Format}
%\bibliography{sample}

\end{document}